\documentclass[onecolumn,preprintnumbers,amsmath,amssymb,aps,prl,superscriptaddress]{revtex4}
\usepackage{graphicx}
\usepackage{textcomp} 
\usepackage{ulem} 
\usepackage[caption=false]{subfig} 
\begin{document}
\title{Elastohydrodynamics of a sliding, spinning and sedimenting cylinder near a soft wall} 
\author{Thomas Salez}
\affiliation{School of Engineering and Applied Sciences, and Department of Physics, Harvard University, Cambridge, MA 02138, USA}
\affiliation{PCT Lab, UMR CNRS 7083 Gulliver, ESPCI ParisTech, PSL Research University, 75005 Paris, France}
\author{L. Mahadevan}
\affiliation{School of Engineering and Applied Sciences, and Department of Physics, Harvard University, Cambridge, MA 02138, USA}

\begin{abstract}
We consider the motion of a fluid-immersed negatively buoyant particle in the vicinity of a thin compressible elastic wall, a situation that arises in a variety of technological and natural settings. We use scaling arguments to establish different regimes of sliding, and complement these estimates using thin-film lubrication dynamics to determine an asymptotic theory for the sedimentation, sliding, and spinning motions of a cylinder. The resulting theory takes the form of three coupled nonlinear singular-differential equations. Numerical integration of the resulting equations confirms our scaling relations and further yields a range of unexpected behaviours. Despite the low-Reynolds feature of the flow, we demonstrate that the particle can spontaneously oscillate when sliding, can generate lift via a Magnus-like effect, can undergo a spin-induced reversal effect, and also shows an unusual sedimentation singularity. Our description also allows us to address a sedimentation-sliding transition that can lead to the particle coasting over very long distances, similar to certain geophysical phenomena. Finally, we show that a small modification of our theory allows to generalize the results to account for additional effects such as wall poroelasticity.
\end{abstract}
\maketitle 

\section{Introduction}
The sedimentation of a heavy solid in a fluid has been studied thoroughly, as the dynamics of settling and sliding are relevant to a broad class of phenomena across many orders of magnitude, ranging from landslides~\cite[]{Campbell1989}, earthquakes~\cite[]{Ma2003}, avalanches~\cite[]{Glenne1987}, to the lubrication of cartilaginous joints~\cite[]{Grodzinsky1978,Mow1984,Mow2002}, and motion of cells in a microfluidic channel~\cite[]{Byun2013} or in a blood vessel~\cite[]{Goldsmith1971}. Following the now classical studies of the dynamics of a particle near a rigid wall~\cite[]{Brenner1962,Goldman1967,Jeffrey1981}, additional effects such as the influence of the boundary conditions~\cite[]{Hocking1973}, and their role on drag~\cite[]{Trahan1985}, viscometry~\cite[]{Wehbeh1993}, and bouncing~\cite[]{Gondret1999}, have been accounted for.  Recently, motion of wedge-like objects down an incline~\cite[]{Cawthorn2010}, as well as the effects of elasticity in such contexts as granular impact~\cite[]{Davis1986}, polymer-bearing contacts~\cite[]{Sekimoto1993}, solvent permeation in gels~\cite[]{Sekimoto1994}, soft lubrication~\cite[]{Skotheim2004,Skotheim2005}, transient effects in displacement-controlled systems~\cite[]{Weekley2006},  settling on soft and poroelastic beds~\cite[]{Balmforth2010,Gopinath2011}, adhesive walls~\cite[]{Mani2012}, and self-similar contact~\cite[]{Snoeijer2013}, have also been addressed. In all these phenomena, the minimal model of motion relates to that of a solid object immersed in a viscous fluid in the vicinity of a soft elastic or poroelastic wall. 

Perhaps surprisingly then, the general theory for the free motion of a rigid solid close to a soft incline -- which has through its degrees of freedom the ability to simultaneously sediment, slide, and spin -- does not seem to have been considered. These modes naturally arise in several applications such as particle capture, joint lubrication, and have analogues in certain geophysical phenomena. Here, we study this problem in a minimal setting and describe the essential scalings and qualitative features, develop a soft lubrication theory that complements these scaling ideas, and solve the resulting equations numerically to characterize the broad range of possible behaviours. 

\section{Mathematical model and scaling analysis}
\begin{figure}
\begin{center}
\includegraphics[width=9cm]{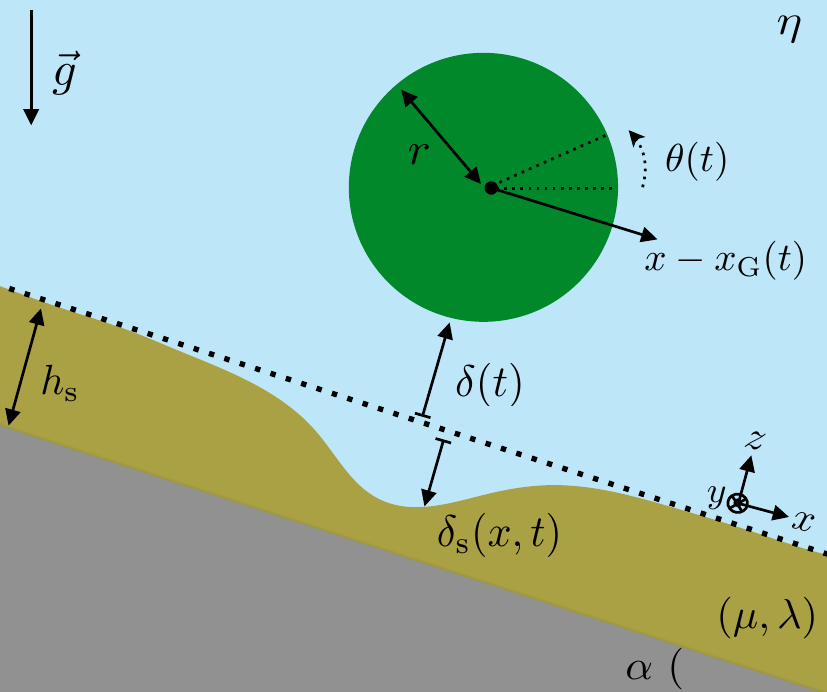}
\end{center}
\caption{\textit{Schematic of the system. A  negatively buoyant cylinder (green) falls down under the acceleration of gravity $\vec{g}$, inside a viscous fluid (blue), in the vicinity of a thin soft wall (brown). The ensemble lies atop a tilted, infinitely rigid support (grey).}}
\label{Fig1}
\end{figure}
We consider the 2D system depicted in~Fig.~\ref{Fig1} that consists of the free gravitational fall of a long cylinder of radius $r$, density $\rho$, mass (per unit length) $m=\pi r^2\rho$, and buoyant mass $m^*=\pi r^2\rho^*=\pi r^2\left(\rho-\rho_{\textrm{fluid}}\right)>0$, where $\rho_{\textrm{fluid}}$ is the density of the neighbouring fluid of viscosity $\eta$. We assume that the motion of the cylinder occurs in the vicinity of a wall that is inclined at an angle $\alpha\in[0,\pi/2]$ with respect to the horizontal direction, and coated with a soft elastic layer of thickness $h_{\textrm{s}}$, and Lam\'e's coefficients $\mu$ and $\lambda$. We denote by $\delta_{\textrm{s}}(x,t)$ the deformation of the fluid-wall interface. Note that a positive indentation of the compressible elastic wall corresponds to a negative value of $\delta_\textrm{s}$. The system is assumed to be invariant along $y$, i.e. we limit ourselves to planar motions wherein the cylinder has three degrees of freedom: the gap $\delta(t)$ between the cylinder and the undeformed wall along $z$, the tangential coordinate $x_{\textrm{G}}(t)$ of the cylinder axis along $x$, and the angle $\theta(t)$ through which the cylinder has rotated.  

We further assume that the cylinder starts its motion at time $t=0$, with $\delta(0)=\delta_0=\epsilon r\ll r$, and possibly non-zero initial translational and angular velocity. Due to this scale separation, we are in the lubrication regime~\cite[]{Batchelor1967}, where the fluid viscous shear stresses are small relative to the flow-induced pressure $p(x,t)$~\footnote{We will add the hydrostatic contribution to the pressure later, through buoyancy of the cylinder. Thus, $p(x,t)$ contains only the flow contribution.}, which itself vanishes far from the contact zone as $x\rightarrow\pm\infty$. The tangential extent $l(t)\gg\delta(t)$ of the flow-induced pressure disturbance scales as $l(t)\sim\sqrt{r\delta(t)}\ll r$, so that as for the Hertzian contact~\cite[]{Johnson1985}, we can assume a parabolic shape of the deformed interface, and the total gap profile may be written as:
\begin{equation}
\label{gapreal}
h(x,t)=\delta(t)-\delta_{\textrm{s}}(x,t)+\frac{\left[x-x_{\textrm{G}}(t)\right]^2}{2r}\ .
\end{equation}

The thin soft compressible wall may also be treated via a lubrication-like theory for elastic deformations if~\footnote{The present study eventually breaks down when $\delta$ is too small, and the condition $h_{\textrm{s}}\ll \sqrt{r\delta}$ is not satisfied. Nevertheless, if for instance $h_{\textrm{s}}\sim\delta_0$, this is a valid assumption since $\delta\sim\delta_0\ll r$.} $h_{\textrm{s}}\ll l(t)$, so that the algebraic displacement $\delta_{\textrm{s}}(x,t)$ of the fluid-wall interface is simply obtained from linear elastic response to the \textit{local} flow-induced pressure disturbance~\cite[]{Skotheim2004,Skotheim2005}:
\begin{equation}
\label{locel}
\delta_{\textrm{s}}(x,t)=-\frac{h_{\textrm{s}}\ p(x,t)}{2\mu+\lambda}\ .
\end{equation}

To characterize the motion of the cylinder near the inclined thin soft wall, we need to calculate the fluid drag force created by the flow-induced pressure field in the contact zone, which is driven by the tangential fluid velocity $u(x,z,t)$ along $x$. We non-dimensionalize the problem using the following choices: $z=Z r\epsilon$, $h=H r\epsilon$, $\delta=\Delta r\epsilon$, $x=Xr\sqrt{2\epsilon}$, $x_{\textrm{G}}=X_{\textrm{G}}r\sqrt{2\epsilon}$, $\theta=\Theta\sqrt{2\epsilon}$, $t=Tr\sqrt{2\epsilon}/c$, $u=Uc$, and $p=P\eta c\sqrt{2}/(r\epsilon^{3/2})$, where we have introduced a free fall velocity scale $c=\sqrt{2gr\rho^*/\rho}$, and the dimensionless parameter:
\begin{equation}
\xi=\frac{3 \sqrt{2}\,\eta}{r^{3/2}\epsilon\sqrt{\rho\rho^*g}}\ .
\end{equation}
This parameter measures the ratio of the free fall time $\sqrt{\rho r\epsilon/(\rho^*g)}$ and the typical lubrication damping time $m\epsilon^{3/2}/\eta$ over which the inertia of the cylinder vanishes. In fact, for a cylinder falling towards a rigid wall, the lubrication drag force (per unit length) exerted in the contact zone reads $\sim-\eta\dot\delta/\epsilon^{3/2}$~\cite[]{Jeffrey1981}. The typical decay time of the cylinder inertia can thus be estimated by balancing this damping force and the inertia (per unit length) $m\ddot{\delta}$, which leads to the above time scale. Note that the power $3/2$ is specific to the 2D case. 

With these definitions, the dimensionless gap profile given by Eq.~(\ref{gapreal}) becomes:
\begin{equation}
\label{gap}
H(X,T)=\Delta(T)+\left[X-X_{\textrm{G}}(T)\right]^2+\kappa P(X,T)\ , 
\end{equation}
where the dimensionless compliance is:
\begin{equation}
\kappa=\frac{2h_{\textrm{s}}\eta\sqrt{g\rho^*}}{r^{3/2}\epsilon^{5/2}(2\mu+\lambda)\sqrt{\rho}}\ .
\end{equation}
To remain in the linearly elastic regime for the wall deformation, we assume that $\kappa \ll 1$. 

Before delving into a detailed theory, we first derive some scaling relations for the sliding dynamics of the cylinder. For steady motions, the analysis of~\cite[]{Sekimoto1993,Skotheim2004,Skotheim2005} shows that one non-trivial effect of the soft substrate is to induce a positive elastohydrodynamic pressure $\sim\eta^2 \dot{x}_{\textrm{G}}^2 rh_{\textrm{s}}/(\mu\delta^{4})$ in the contact zone, when the particle is translated uniformly along the wall at speed $\dot{x}_{\textrm{G}}$ while being at a constant distance $\delta$. In the present 2D-like case of a free cylinder, when that pressure is integrated once along the contact length $l\sim\sqrt{r\delta}$, this leads to a net positive elastohydrodynamic lift force (per unit length) $\sim\eta^2 \dot{x}_{\textrm{G}}^2 r^{3/2}h_{\textrm{s}}/(\mu\delta^{7/2})$ that tends to repel the sliding particle away from the soft wall. Since the force of gravity (per unit length)  $\sim\rho^*gr^2\cos\alpha$ tends to bring it back towards the wall, balancing the two forces allows one to predict a  sliding height as a function of the speed $\dot{x}_{\textrm{G}}$, given by:
\begin{equation}
\label{deltaeq}
\delta_{\textrm{eq}}\sim\left(\frac{h_{\textrm{s}}\eta^2\dot{x}_{\textrm{G}}^2}{\mu\sqrt{r}\rho^*g\cos\alpha}\right)^{2/7}\ .
\end{equation}

When the sliding velocity $\dot{x}_{\textrm{G}}$ does not vary much -- as is often the case on short time scales when damping  in the normal direction  $z$ is much stronger than in the tangential direction $x$ -- this represents a stable equilibrium gap thickness. A small perturbation about this equilibrium position suggests that the the cylinder will oscillate with frequency $\sim\sqrt{\rho^*g\cos\alpha/(\rho\delta_{\textrm{eq}})}$, even as the lubrication viscous damping will cause these inertial oscillations to decay over a typical time $\sim (\delta_{\textrm{eq}}/r)^{3/2}m/\eta$, as already introduced above. 

Finally, after a transient evolution along the tilted wall, we expect the cylinder to reach a long-term steady-state sliding regime characterized by a terminal velocity $u_{\infty}$ and a constant gap thickness $\delta_{\infty}$. Leaving aside the conditions of existence of this scenario for now, we can already describe the properties of this regime using simple arguments. Along $z$, the gravity-vs-lift force balance leads to Eq.~(\ref{deltaeq}) above, with $\dot{x}_{\textrm{G}}=u_{\infty}$ and $\delta_{\textrm{eq}}=\delta_{\infty}$. The second equation we need comes from the power balance in the direction of sliding motion $x$. The power (per unit length) $\sim u_{\infty}\rho^*gr^2\sin\alpha$ generated by the gravitational driving is entirely dissipated in the contact zone through the viscous damping power $\sim\eta(u_{\infty}/\delta_{\infty})^2l\delta_{\infty}\sim\eta u_{\infty}^2\sqrt{r/\delta_{\infty}}$. This leads to the expressions for the steady gap and terminal velocity, given by:
\begin{eqnarray}
\label{scaling1}
\delta_{\infty}&\sim&\frac{\rho^{*\,2/5}g^{2/5}rh_{\textrm{s}}^{2/5}\sin^{4/5}\alpha}{\mu^{2/5}\cos^{2/5}\alpha}\ ,\\
u_{\infty}&\sim&\frac{\rho^{*\,6/5}g^{6/5}r^2h_{\textrm{s}}^{1/5}\sin^{7/5}\alpha}{\eta\mu^{1/5}\cos^{1/5}\alpha}\nonumber\ .
\end{eqnarray}

With these scaling relationships in place, we now aim at constructing a detailed soft lubrication theory that goes beyond these arguments and, as we will see, introduces new phenomena as well.

\section{Soft lubrication theory}
In the thin-gap limit, the governing Stokes equations for incompressible viscous flow are given in scaled form by~\cite[]{Reynolds1886,Batchelor1967,Oron1997}:
\begin{equation}
\label{stokes}
U_{ZZ}=P_X\ ,
\end{equation}
together with no-slip boundary conditions, $U(X,Z=-\kappa P,T)=0$ and $U(X,Z=H-\kappa P,T)=\dot{X}_{\textrm{G}}+\dot{\Theta}$. Solving Eq.~(\ref{stokes}) with the above boundary conditions, and invoking the condition of volume conservation:
\begin{equation}
\label{tfe}
\partial_TH+\partial_X\int_{-\kappa P}^{H-\kappa P} dZ\ U=0\ ,
\end{equation}
yields the following equation for the evolution of the gap:
\begin{equation} 
\label{eqgen3}
12\dot{\Delta}-24\left(X-X_{\textrm{G}}\right)\dot{X}_{\textrm{G}}+12\kappa P_T=\left[H^3P_X-6\left(\dot{X}_{\textrm{G}}+\dot{\Theta}\right)H\right]_X\ .
\end{equation}
\begin{figure}
\begin{center}
\includegraphics[width=13.5cm]{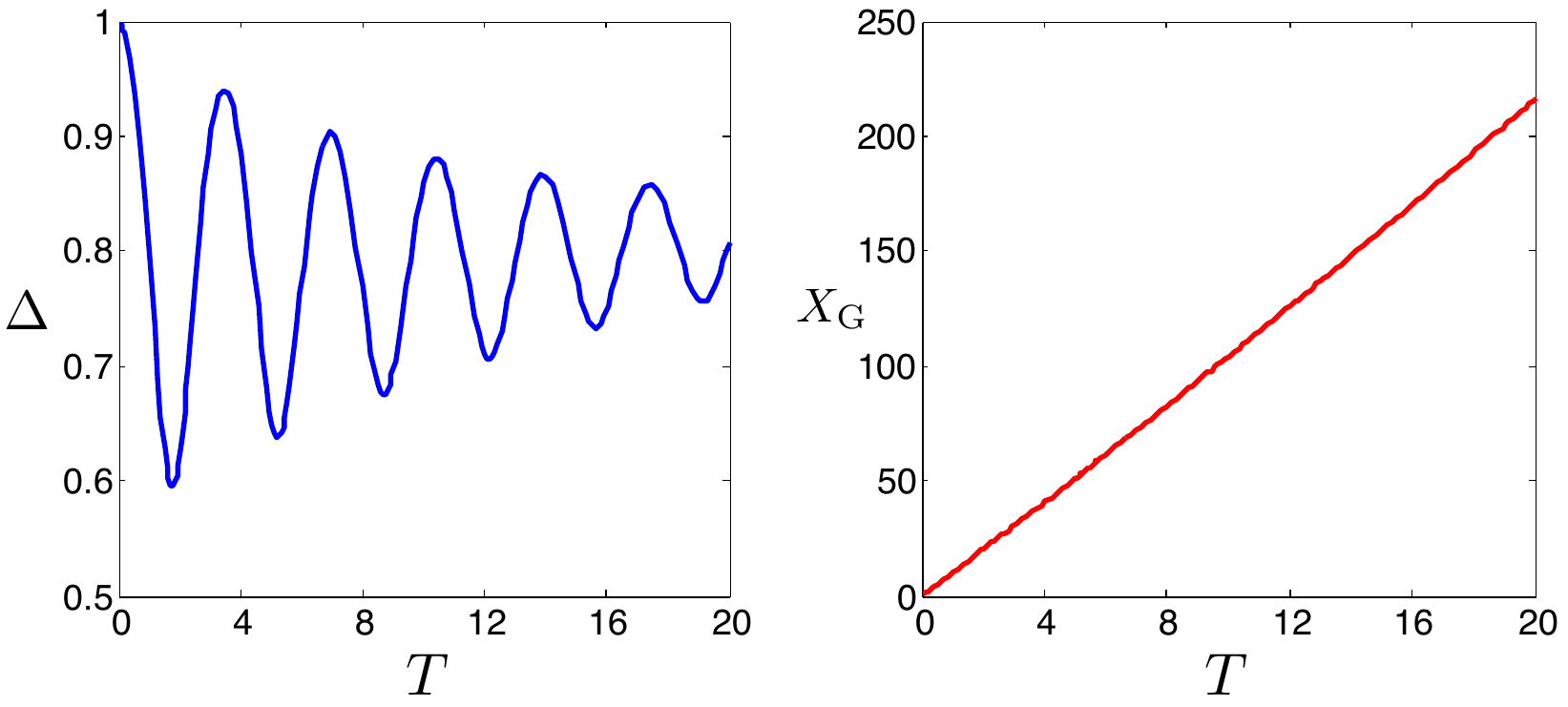}
\end{center}
\caption{\textit{Oscillations. When a cylinder is released close to an inclined wall with a nonzero tangential velocity, it spontaneously oscillates about the stable sliding height; however with time, these oscillations eventually decay. These results follow from the numerical solution of~Eq.~(\ref{tdirect}), for $\alpha=\pi/4$, $\xi=0.1$, $\kappa=0.1$, $\epsilon=0.1$, $\Delta(0)=1$, $X_{\textrm{G}}(0)=\dot{\Delta}(0)=\dot{\Theta}(0)=0$, and $\dot{X}_{\textrm{G}}(0)=10$.}}
\label{Fig2}
\end{figure}

The solution of this equation allows us to evaluate the total pressure-induced drag exerted on the cylinder, through: 
\begin{equation}
\label{dragp}
\vec{D_p}\approx\int_{-\infty}^{\infty}dX\ P\ \vec{e_{z}}-\sqrt{2\epsilon}\int_{-\infty}^{\infty}dX \left(X-X_{\textrm{G}}\right) P\ \vec{e_{x}}\ ,
\end{equation}
where we have used the fact that the normal vector to the cylinder surface {is} $\vec{n}\approx\vec{e_{z}}-\frac{x-x_{\textrm{G}}}{r}\vec{e_{x}}$. Similarly, the shear drag force exerted on the cylinder along the $\vec{e_{x}}$ axis is given by:
\begin{equation}
\label{drags}
D_{\sigma,\parallel}=-\sqrt{\frac{\epsilon}{2}}\int_{-\infty}^{\infty}dX\ U_Z|_{Z=H-\kappa P}\ .
\end{equation}

When the dimensionless compliance is assumed to satisfy $\kappa \ll 1$, we may employ perturbation theory in this parameter~\cite[]{Skotheim2004,Skotheim2005}, using the following expansion for the pressure, $P=P^{(0)}+\kappa P^{(1)}$, where $P^{(0)}|_{X\rightarrow\pm \infty} =P^{(1)}|_{X\rightarrow\pm \infty} = 0$. As detailed in Appendix A and B, integrating Eq.~(\ref{eqgen3}) to first order in $\kappa$, and using Eqs.~(\ref{dragp}) and~(\ref{drags}), leads to the following expressions for the perpendicular drag along $\vec{e_{z}}$ and the two parallel components along $\vec{e_{x}}$:
\begin{eqnarray}
\label{3drag}
D_{p,\perp}&=&-\frac{3\pi}{2}\frac{\dot{\Delta}}{\Delta^{3/2}}+\kappa\left[\frac{45\pi\ddot{\Delta}}{16\Delta^{7/2}}-\frac{63\pi\dot{\Delta}^2}{8\Delta^{9/2}}+\frac{3\pi\left(\dot{\Theta}-\dot{X}_{\textrm{G}}\right)^2}{8\Delta^{7/2}}\right]\ ,\\
D_{p,\parallel}&=&\pi\sqrt{2\epsilon}\ \frac{\dot{\Theta}-\dot{X}_{\textrm{G}}}{\sqrt{\Delta}}+\kappa\sqrt{\frac{\epsilon}{2}}\left[ \frac{23\pi\dot{\Delta}\left(\dot{\Theta}-\dot{X}_{\textrm{G}}\right)}{8\Delta^{7/2}}+\frac{\pi\left(\ddot{X}_{\textrm{G}}-\ddot{\Theta}\right)}{2\Delta^{5/2}}\right]\ ,\nonumber\\
D_{\sigma,\parallel}&=&-\pi\sqrt{2\epsilon}\frac{\dot{\Theta}}{\sqrt{\Delta}}+\kappa\sqrt{\frac{\epsilon}{2}}\left[\frac{\pi\left(\ddot{\Theta}-\ddot{X}_{\textrm{G}}\right)}{4\Delta^{5/2}}+\frac{\pi\dot{\Delta}\dot{X}_{\textrm{G}}}{2\Delta^{7/2}}-\frac{19\pi\dot{\Delta}\dot{\Theta}}{8\Delta^{7/2}}\right]\nonumber\ .
\end{eqnarray}

We stress that we have neglected the forces acting outside the contact zone, consistent with the lubrication approximation. To justify this choice, let us first consider the sedimentation motion towards the rigid wall. The drag force (per unit length) exerted on a cylinder in a bulk fluid scales as $d_{\textrm{bulk}}\sim\eta\dot\delta$~\cite[]{Brenner1962}. According to Eq.~(\ref{3drag}), the pressure-induced lubrication drag force (per unit length) reads, in real variables, $d_{p,\perp}=2c\eta D_{p,\perp}/\epsilon\sim\eta\dot{\delta}(r/\delta)^{3/2}$~\cite[]{Jeffrey1981}.  Since $\delta\ll r$ in the lubrication approximation, one can safely neglect the bulk drag against the lubrication one acting in the contact zone. Similarly, according to Eq.~(\ref{3drag}), for the tangential motion along a rigid wall, the pressure-induced drag scales as $d_{p,\parallel}=2c\eta D_{p,\parallel}/\epsilon\sim \eta\dot{x}_{\textrm{G}}\sqrt{r/\delta}$~\cite[]{Jeffrey1981}, which -- despite being smaller than $d_{p,\perp}$ -- is once again larger than $d_{\textrm{bulk}}$, in the lubrication approximation. One can thus safely neglect the bulk drag relative to the lubrication drag for the tangential degree of freedom as well. Since the shear-induced drag is of the same order and symmetry as the tangential pressure-induced drag, the previous conclusion extends immediately to the rotational degree of freedom. We note that the argument above assumes a rigid wall, since the fluid lubrication order is not modified by the softness of the wall. As an illustration of this statement, all the softness-induced terms -- the ones proportional to the independent compliance parameter $\kappa$ in Eq.~(\ref{3drag}) -- have the same order in $\epsilon$ as the corresponding terms for the rigid wall.

We also note that it may not be satisfactory at first sight to get an acceleration-dependent drag, even as a first order correction, as it means at time $T=0$, when there is no flow, there is a pressure field that deforms the wall. To understand this, we note that the origin of this behaviour is to be found in the $P_T$ term in Eq.~(\ref{eqgen3}), since $P\propto\dot{\Delta}$ due to the Stokes equation. In our analysis, we have neglected the linearized inertia of the fluid, $\rho_{\textrm{fluid}}\partial_tu$, but at very short times this term becomes dominant and resolves this apparent paradox.
\begin{figure}
\begin{center}
\includegraphics[width=13.5cm]{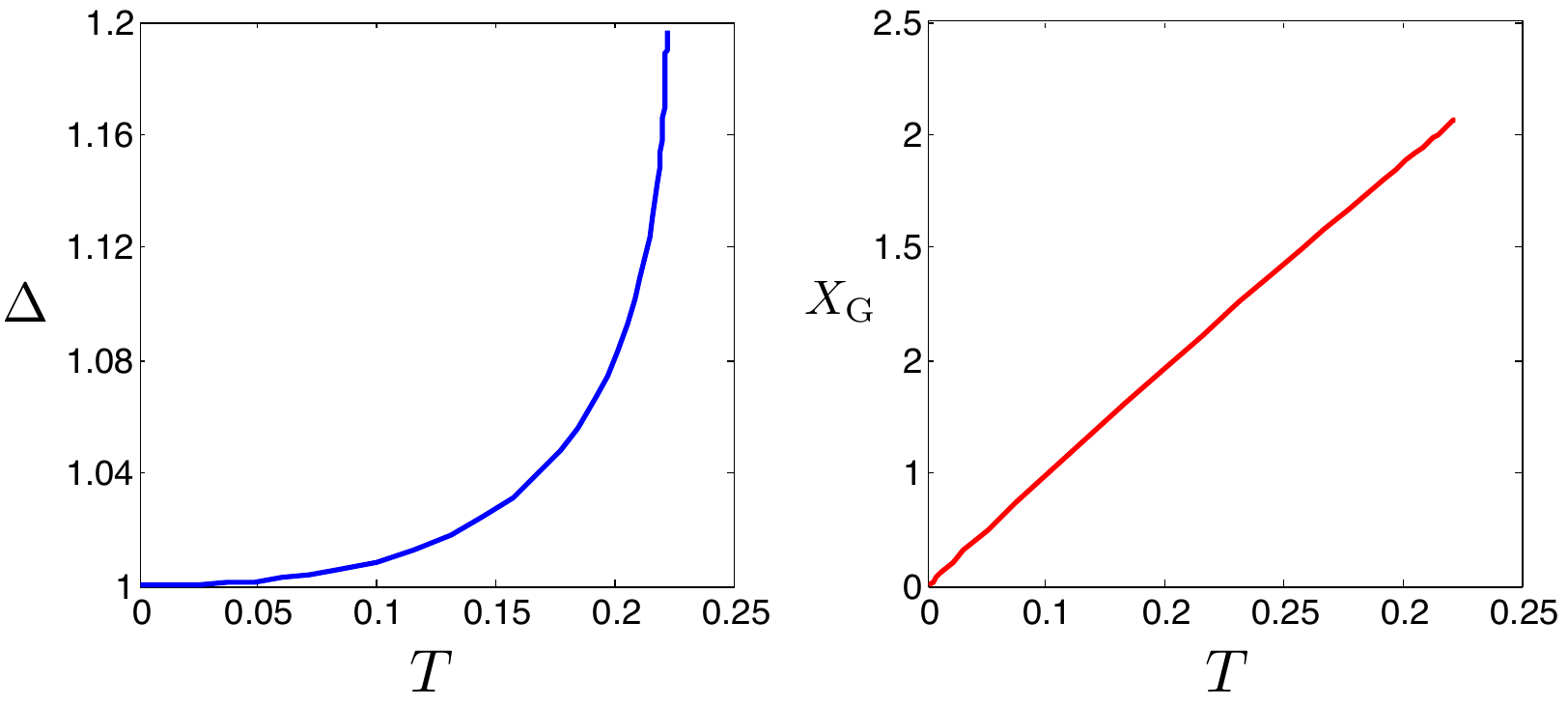}
\end{center}
\caption{\textit{Magnus-like effect. When the cylinder is released close to the horizontal elastic wall with a nonzero angular velocity, it lifts off. These results follow from the numerical solution of~Eq.~(\ref{tdirect}), for $\alpha=0$, $\xi=10$, $\kappa=0.1$, $\epsilon=0.1$, $\Delta(0)=1$, $X_{\textrm{G}}(0)=\dot{\Delta}(0)=0$, and $\dot{X}_{\textrm{G}}(0)=\dot{\Theta}(0)=10$. If we replace  the last condition by $\dot{\Theta}(0)=0$, then $\Delta$ diminishes.}}
\label{Fig3}
\end{figure}

Knowing the dominant elastohydrodynamic drag forces acting on the cylinder, we now use the balance of linear and angular momentum (see Appendix A and B) to write down the coupled nonlinear differential equations for the translational and rotational motions of the cylinder, as it sediments, slides and rolls down the incline:
\begin{eqnarray}
\ddot{\Delta}&=&-\xi\frac{\dot{\Delta}}{\Delta^{3/2}}-\frac{\kappa\xi}{4}\left[21\frac{\dot{\Delta}^2}{\Delta^{9/2}}-\frac{\left(\dot{\Theta}-\dot{X}_{\textrm{G}}\right)^2}{\Delta^{7/2}}-\frac{15}{2}\frac{\ddot{\Delta}}{\Delta^{7/2}}\right]-\cos\alpha\ ,\nonumber\\ 
\ddot{X}_{\textrm{G}}&=&-\frac{2\epsilon\xi}{3}\ \frac{\dot{X}_{\textrm{G}}}{\sqrt{\Delta}}-\frac{\kappa\epsilon\xi}{6}\left[\frac{19}{4}\frac{\dot{\Delta}\dot{X}_{\textrm{G}}}{\Delta^{7/2}}-\frac{\dot{\Delta}\dot{\Theta}}{\Delta^{7/2}}+\frac12\frac{\ddot{\Theta}-\ddot{X}_{\textrm{G}}}{\Delta^{5/2}}\right]+\sqrt{\frac{\epsilon}{2}}\sin\alpha\ ,\nonumber\\
\label{tdirect}
\ddot{\Theta}&=&-\frac{4\epsilon\xi}{3}\frac{\dot{\Theta}}{\sqrt{\Delta}}-\frac{\kappa\epsilon\xi}{3}\left[\frac{19}{4}\frac{\dot{\Delta}\dot{\Theta}}{\Delta^{7/2}}-\frac{\dot{\Delta}\dot{X}_{\textrm{G}}}{\Delta^{7/2}}+\frac12\frac{\ddot{X}_{\textrm{G}}-\ddot{\Theta}}{\Delta^{5/2}}\right]\ .
\end{eqnarray}
We note that the lubrication pressure-induced torque vanishes since the corresponding forces act along the radii of the cylinder. Interestingly, this would not be the case for the opposite case of a soft cylinder -- which will deform asymmetrically -- near a rigid wall, thus breaking once a well admitted symmetry between the two dual systems in elasticity~\cite[]{Johnson1985}.

We see that particle inertia plays a central role in Eq.~(\ref{tdirect}), even though we have neglected fluid inertia. To justify this assumption, let us consider for instance a $x$-translation of the cylinder along the rigid wall, at typical speed $c$ and distance $\delta_0$ from the wall. In the Navier-Stokes equation, the local fluid inertia term reads $\rho_{\textrm{fluid}}\partial_tu\sim\rho_{\textrm{fluid}}c/\tau$, where $\tau\sim l/c$ is the typical time scale of the flow at speed $c$, and $l\sim\sqrt{r\delta_0}$ the length of the contact zone along $x$. Similarly, the convective inertia term reads $\rho_{\textrm{fluid}}u\partial_xu\sim\rho_{\textrm{fluid}}c^2/l\sim\rho_{\textrm{fluid}}c/\tau$. On the other hand, the viscous term in the Navier-Stokes equation reads $\eta\partial^2_{zz}u\sim\eta c/\delta_0^2$. The ratio of inertia over viscous terms thus reads $\sim\textrm{Re}\,\epsilon\sim\textrm{Re}\,\delta_0/r$, where the Reynolds number is given by $\textrm{Re}=\rho_{\textrm{fluid}} lc/\eta$.  As for particle inertia, following Newton's law, we note that it scales as $\sim \rho r^2l/\tau^2$. According to our Eq.~(\ref{3drag}), for the tangential motion along the rigid wall, the pressure-induced force (per unit length) scales as $d_{p,\parallel}=2c\eta D_{p,\parallel}/\epsilon\sim\eta c/\sqrt{\epsilon}$~\cite[]{Jeffrey1981}, in real variables. The ratio of particle inertia and fluid viscosity thus reads $\sim(\rho/\rho_{\textrm{fluid}})\textrm{Re}/\sqrt{\epsilon}$, which is much larger than the ratio of fluid inertia and fluid viscosity -- due to the lubrication parameter $\epsilon\ll1$ -- even in the case when the densities are matched. Thus, we see that even if fluid inertia plays a role on short time scales, there is a range of parameters over which it is negligible while particle inertia is still important. This  conclusion remains valid even in the presence of the additional compliance parameter $\kappa$ describing the wall softness.

\section{Behaviour of solutions}
The elastohydrodynamic drag terms on the right-hand sides of Eq.~(\ref{tdirect}) trigger an interesting zoology of solutions, which we now turn to. The solutions are governed by four dimensionless control parameters corresponding to a ratio $\xi$ of viscous damping over gravitational driving, an incline angle $\alpha$, a scaled wall compliance $\kappa\ll1$, and a scaled lubrication gap $\epsilon\ll1$. In addition, there are three relevant initial conditions: $\dot{\Delta}(0)$, $\dot{X}_{\textrm{G}}(0)$, and $\dot{\Theta}(0)$, since $\Delta(0)=1$ by virtue of our choice of the dimensionless variable $\Delta=\delta/\delta_0$, while all the initial tangential positions $X_{\textrm{G}}(0)$ and initial angles $\Theta(0)$ are equivalent. Below, we give a brief flavour of some of the unexpected behaviours of the system with the aim of sketching the diversity of solutions, potentially valid for a variety of similar systems and experiments, rather than to build a complete phase-diagram for this 2D case. 

\subsection{Zoology}
\begin{figure}
\begin{center}
\includegraphics[width=13.5cm]{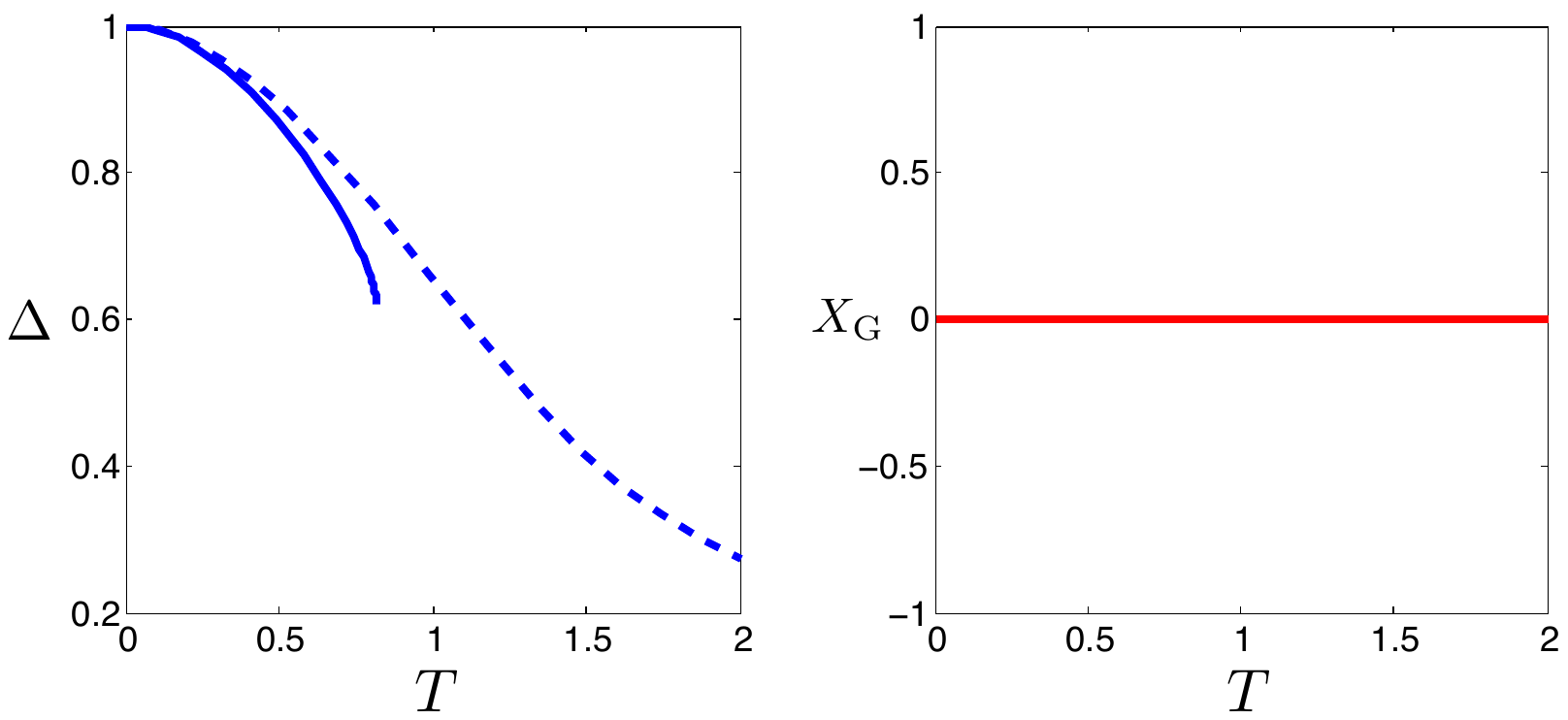}
\end{center}
\caption{\textit{Enhanced sedimentation. When the cylinder is allowed to fall freely vertically towards the horizontal elastic wall, it sediments faster than if the wall is rigid. These results follow from the numerical solutions of~Eq.~(\ref{tdirect}), for $\alpha=0$, $\xi=1$, $\epsilon=0.1$, $\Delta(0)=1$, $X_{\textrm{G}}(0)=\dot{\Delta}(0)=\dot{X}_{\textrm{G}}(0)=\dot{\Theta}(0)=0$. The dashed line corresponds to the case of a rigid wall, with $\kappa=0$, showing that $\Delta$ decreases gradually, while the solid line corresponds to sedimentation towards a soft wall, with $\kappa=0.1$, where the cylinder abruptly crashes downwards after sedimenting at a rate faster than towards a rigid wall. The figure on the right side shows that the horizontal position of the cylinder does not vary at all during sedimentation (the simulation is extended to $T=2$ corresponding to the case of sedimentation towards a rigid wall).}}
\label{Fig4}
\end{figure}
In Fig.~\ref{Fig2}, we show that when the cylinder is released along a steep incline, it slides along it uniformly even as it spontaneously oscillates, although these oscillations are damped. Indeed, the envelope decays over a dimensionless time that is consistent with our earlier scaling estimate: $\Delta_{\textrm{eq}}^{3/2}/\xi\sim6.4$ (in dimensionless form) for the parameters of Fig.~\ref{Fig2}. Similarly, the equilibrium height  can be calculated by balancing gravity $\cos\alpha$ and the elastohydrodynamic lift $\kappa\xi\dot{X}_{\textrm{G}}^2/(4\Delta^{7/2})$ in the first line of Eq.~(\ref{tdirect}), to yield:
\begin{equation}
\label{splan}
\Delta_{\textrm{eq}}=\frac{1}{2^{4/7}}\left(\frac{\kappa\xi\dot{X}_{\textrm{G}}^2}{\cos\alpha}\right)^{2/7}\ ,
\end{equation}
consistent with the dimensional scaling form given in Eq.~(\ref{deltaeq})). For the parameters in Fig.~\ref{Fig2}, one obtains $\Delta_{\textrm{eq}}\approx0.74$, which is close to the observed average value of $\sim0.79$ seen in Fig.~\ref{Fig2}a; the slight difference is due to the weak influence of other terms in Eq.~(\ref{tdirect}). 

In Fig.~\ref{Fig3}, we show another peculiar effect associated with the case when the cylinder is started with an initial spin. As seen, it can lift off the soft wall via a Magnus-like effect~\cite[]{Dupeux2011} even as it slides along a horizontal wall. This effect is due to the fluid shear induced by rotation that leads to an increased hydrodynamic pressure, which deforms the wall and thence leads to a normal force.
\begin{figure}
\begin{center}     
\includegraphics[width=13.5cm]{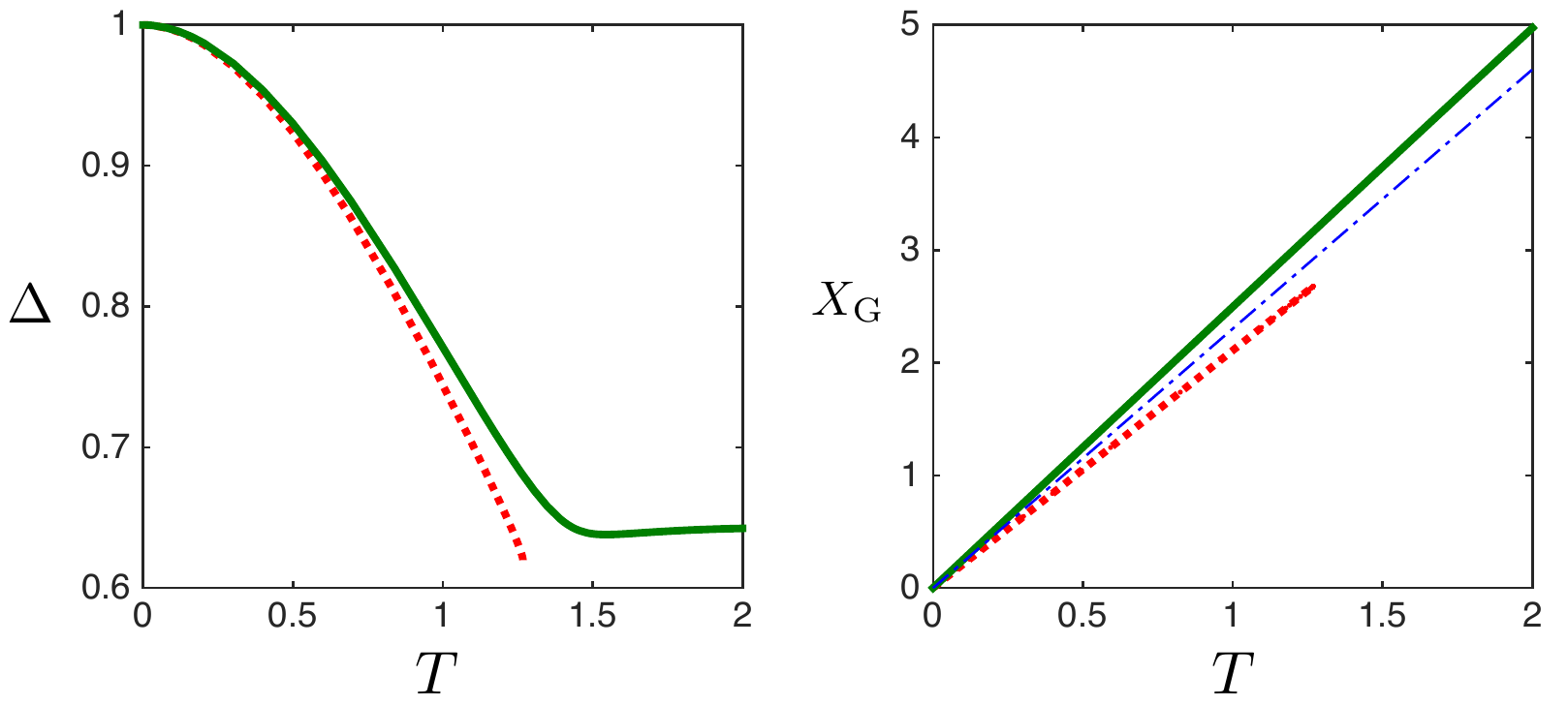}  
\end{center}
\caption{\textit{Sedimentation-sliding transition, observed when   the cylinder is released close to the inclined elastic wall with the threshold tangential velocity given by Eq.~(\ref{tresh}). The results follow from the numerical solution of~Eq.~(\ref{tdirect}), for $\alpha=\pi/4$, $\xi=1$, $\kappa=0.1$, $\epsilon=0.1$, $\Delta(0)=1$, $X_{\textrm{G}}(0)=\dot{\Delta}(0)=\dot{\Theta}(0)=0$, $\dot{X}_{\textrm{G}}(0)=2.1$ (red dots), and $\dot{X}_{\textrm{G}}(0)=2.5$ (green lines). For these parameters, $U_{\textrm{c}}\approx2.30$ (blue dash-dotted line), as obtained from Eq.~(\ref{tresh}).}}
\label{Fig5}
\end{figure} 

Next, we turn to examine the equations when the effective mass of the particle vanishes. Indeed, since we kept both the cylinder inertia and the acceleration drag, as explained above, those two second-derivative terms may cancel each other. This singularity leads to a vanishing effective mass and thus a diverging acceleration $\ddot{\Delta}$, and occurs at the three critical heights:
\begin{eqnarray}
\label{crit}
\Delta_{\textrm{c}1}&=&\left(\frac{15\kappa\xi}{8}\right)^{2/7}\ ,\\
\Delta_{\textrm{c}2}&=&\left(\frac{\epsilon\kappa\xi}{12}\right)^{2/5}\ ,\\
\Delta_{\textrm{c}3}&=&\left(\frac{\epsilon\kappa\xi}{6}\right)^{2/5}\ .
\end{eqnarray}
In Fig.~\ref{Fig4}, we show the evolution of the height when a relatively heavy cylinder is released above a horizontal soft wall. We see that the particle sediments~\footnote{$\Delta$ is not exactly a cylinder-wall distance, since there is an additional $\kappa P(X_{\textrm{G}},T)$ term according to Eq.~(\ref{gap}). However, this additional term can be computed from Eq.~(\ref{tiltpres}) and the solutions of~Eq.~(\ref{tdirect}), if needed.} at an enhanced rate relative to the case when the soft wall is replaced by a rigid wall, corresponding to $\kappa=0$. This is due to the fact that when the cylinder reaches the largest critical height $\Delta_{\textrm{c}1}<1$ of Eq.~(\ref{crit}), the vanishing effective inertial mass leads to an infinite acceleration. This unphysical effect is regularized when one takes into account fluid inertia, leading to a smoothed out temporal profile of sedimentation. Nonetheless, as explained above, the temporal cut-off associated with fluid inertia is assumed to correspond to time scales smaller than the ones associated with the motion of the cylinder, which means that, even if finite, $\ddot{\Delta}$ may be large and the behaviour still very sharp. Finally, we note that when $\Delta_{\textrm{c}1}>1$ and $\dot{\Delta}<0$, the singularity may instead occur at $\Delta_{\textrm{c}3}>\Delta_{\textrm{c}2}$.

We  now use our results to characterize the sedimentation-sliding transition for a cylinder falling down an incline. Equation~(\ref{splan}) suggests that the cylinder can stably slide at a dimensionless height $\Delta_{\textrm{eq}}$. On the other hand, if for instance $\Delta_{\textrm{c}1}<1$ in Eq.~(\ref{crit}), $\Delta_{\textrm{c}1}$ fixes the relevant singular sedimentation height that may be encountered during the fall of the cylinder, as illustrated in Fig.~\ref{Fig4}. The balance of these two dimensionless heights  yields the threshold tangential velocity $U_{\textrm{c}}$ above which sliding becomes possible:
\begin{equation}
\label{tresh}
U_{\textrm{c}}=\sqrt{\frac{15}{2}}\ \sqrt{\cos\alpha}\ .
\end{equation}
In fact, if $\Delta_{\textrm{eq}}<\Delta_{\textrm{c}1}<1$, and thus $\dot{X}_{\textrm{G}}<U_{\textrm{c}}$, the singular sedimentation height $\Delta_{\textrm{c}1}$ is reached before the sliding height $\Delta_{\textrm{eq}}$, and one typically gets sedimentation. If, in contrast, $\Delta_{\textrm{eq}}>\Delta_{\textrm{c}1}$, and thus $\dot{X}_{\textrm{G}}>U_{\textrm{c}}$, the sliding height $\Delta_{\textrm{eq}}$ is reached before the singular sedimentation height $\Delta_{\textrm{c}1}$, and one typically gets sliding. 
This transition is illustrated in Fig.~\ref{Fig5}, for two given sets of dimensionless parameters and initial conditions. For instance, with a meter-sized body, this reasonably corresponds to a $\sim1\ \textrm{m}/\textrm{s}$ threshold velocity. We note that, although the presence of an elastic wall is crucial in the underlying mechanism, the elastohydrodynamic details do not appear in this purely gravitational expression.  
\begin{figure}
\begin{center}
\includegraphics[width=13.5cm]{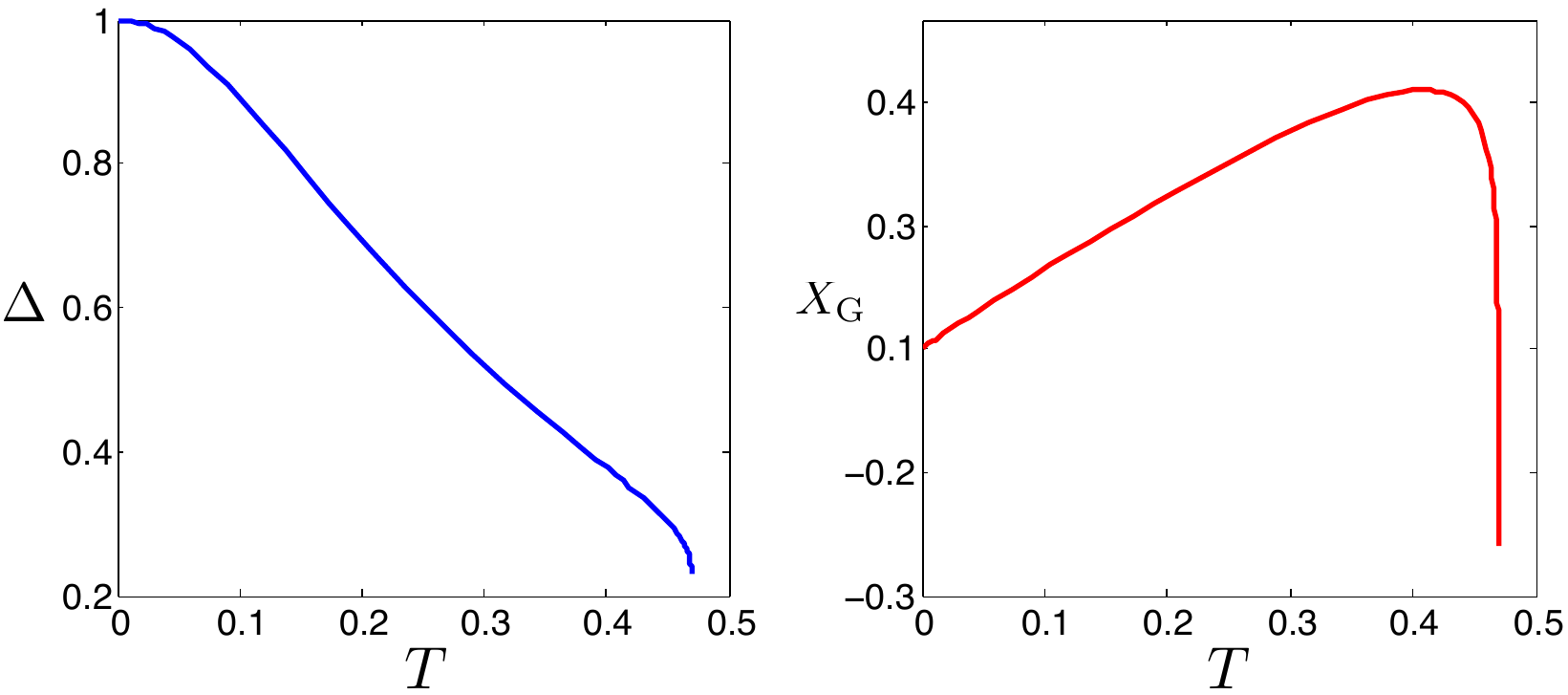}
\end{center}
\caption{\textit{Spin-induced reversal. When the cylinder is released close to the horizontal elastic wall with nonzero tangential and rotational velocities, it can return backwards. These results follow from the numerical solution of~Eq.~(\ref{tdirect}), for $\alpha=0$, $\xi=10$, $\kappa=0.1$, $\epsilon=0.1$, $\Delta(0)=1$, $X_{\textrm{G}}(0)=\dot{\Delta}(0)=0$, $\dot{X}_{\textrm{G}}(0)=1$, and $\dot{\Theta}(0)=10$.}}
\label{Fig6}
\end{figure}

We conclude our tour of the zoology of solutions by noting that when a relatively heavy cylinder is released with spin and tangential velocity, it can reverse its direction of motion and return backwards along the soft wall, as shown in Fig.~\ref{Fig6}. This effect can be understood by noting that the second equation in Eq.~(\ref{tdirect}) characterizes the dynamics of sliding. Thus, when $\dot{\Delta}<0$, a large enough positive spin velocity suffices to bring about a reversal in the tangential acceleration.

\subsection{Long-term steady sliding}
\begin{figure}
\begin{center}     
\includegraphics[width=13.5cm]{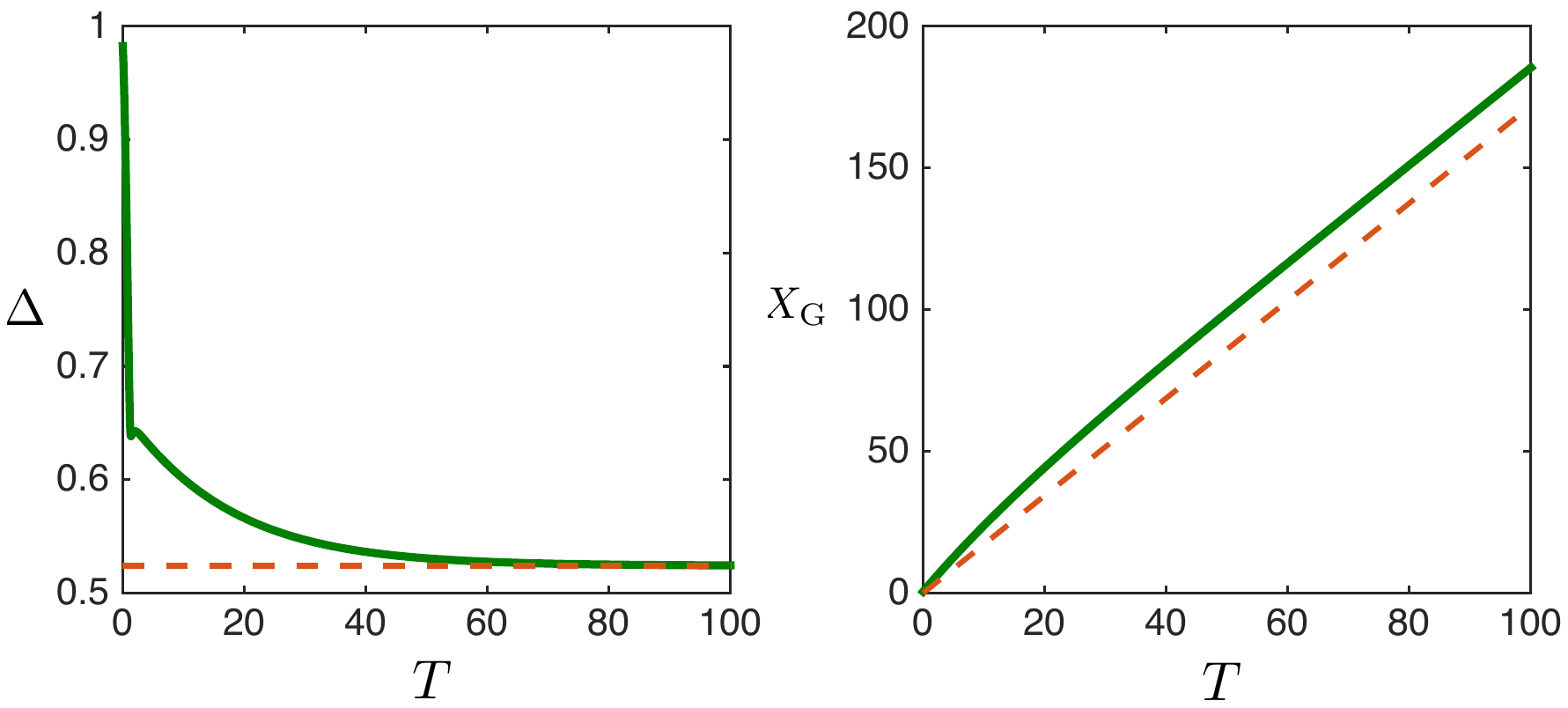}  
\end{center}
\caption{\textit{Convergence to the long-term sliding steady-state, observed when the cylinder is released close to the inclined elastic wall with an initial tangential velocity greater than the threshold velocity given by Eq.~(\ref{tresh}). The results follow from the numerical solution of~Eq.~(\ref{tdirect}), for $\alpha=\pi/4$, $\xi=1$, $\kappa=0.1$, $\epsilon=0.1$, $\Delta(0)=1$, $X_{\textrm{G}}(0)=\dot{\Delta}(0)=\dot{\Theta}(0)=0$, and $\dot{X}_{\textrm{G}}(0)=2.5$ (green lines). For those parameters, the terminal height and velocity of the sliding steady-state are given by $\Delta_{\infty}\approx0.524$ and $U_{\infty}\approx1.72$ (orange dashed lines), as obtained from Eqs.~(\ref{ssv}) and~(\ref{ssh}) respectively.}}
\label{Fig7}
\end{figure}

Once initiated and stabilized, the sliding motion eventually reaches a long-term steady-state, with a terminal velocity that reads:
\begin{equation}
\label{ssv}
U_{\infty}=\frac{3^{7/5}}{2^{5/2}}\ \frac{ \kappa^{1/5}\sin^{7/5}\alpha}{\xi^{6/5}\epsilon^{7/10}\cos^{1/5}\alpha}\ ,
\end{equation}
which is obtained by balancing viscous damping and gravity in the tangential component of Eq.~(\ref{tdirect}), and by replacing $\dot{X}_{\textrm{G}}$ and $\Delta_{\textrm{eq}}$ with $U_{\infty}$ and $\Delta_{\infty}$ in Eq.~(\ref{splan}), respectively. This also leads to a prediction of the associated terminal sliding height:
\begin{equation}
\label{ssh}
\Delta_{\infty}=\frac{3^{4/5}}{4}\ \frac{ \kappa^{2/5}\sin^{4/5}\alpha}{\xi^{2/5}\epsilon^{2/5}\cos^{2/5}\alpha}\ ,
\end{equation}
consistent with the scaling relations in Eq.~(\ref{scaling1}). The convergence to this long-term steady-state for the stable sliding case is illustrated by solving Eq.~(\ref{tdirect}) and the results are depicted in Fig.~\ref{Fig7}, showing that the cylinder indeed reaches the terminal velocity and height obtained above.

Naturally, these results are valid as long as $\Delta$ remains sufficiently smaller than $\sim\epsilon^{-1}$, so that the lubrication approximation holds. This criterion corresponds to a terminal velocity $U_{\infty}$ being smaller than $\sim\sqrt{\cos\alpha/\left(\kappa\xi\epsilon^{7/2}\right)}$.

\section{Role of poroelasticity}
We conclude with a brief discussion of a generalisation of our results to the case when the wall is fluid permeable, a problem of some relevance to many biological and geological situations~\cite[]{Biot1941,Burridge1981,Gopinath2011}, and we follow and generalize the results of ~\cite[]{Skotheim2004b,Skotheim2005,Gopinath2011} that we summarize below.

We introduce the volume fraction $\phi$ of fluid in the porous wall, the bulk modulus $\beta^{-1}\gg \mu$ of the solid porous matrix (with $\mu$ now being the composite shear modulus of the poroelastic medium), and the isotropic Darcy permeability $k$, and we note that the pore size $\sim\sqrt{k}$ is small in comparison with the wall thickness $h_{\textrm{s}}$. We assume that there is no flow inside the poroelastic wall in comparison with the flow in the lubrication gap, which is valid as long as $kh_{\textrm{s}}\ll\delta(t)^3$. For example, if $h_{\textrm{s}}\sim\delta_0$, this follows due to scale separation.

The fluid-permeable wall introduces a new time scale associated with flow-induced stress relaxation given by $\tau_{\textrm{p}}\sim\eta h_{\textrm{s}}^2/(k\mu)$, which has to be compared with the lubrication time scale $\tau\sim r\sqrt{\epsilon}/c$~\cite[]{Skotheim2005}. If $\tau\gg\tau_{\textrm{p}}$, the fluid in the wall is in equilibrium with the outside and a purely elastic theory suffices, so that Eq.~(\ref{locel}) is modified to read:
\begin{equation}
\delta_{\textrm{s}}(x,t)=-\frac{h_{\textrm{s}}(1-\phi)}{2\mu+\lambda}\,p(x,t)\ ,
\end{equation}
which simply corresponds to a small effective stiffening due to the presence of a volume fraction $\phi$ of fluid in the poroelastic wall. In contrast, if $\tau\ll\tau_{\textrm{p}}$, the pore fluid has no time to adapt and we find that the wall is effectively stiffer, with $(2\mu+\lambda)\rightarrow \phi/\beta$, so that:
\begin{equation}
\delta_{\textrm{s}}(x,t)=-\frac{\beta h_{\textrm{s}}}{\phi}\ p(x,t)\ .
\end{equation}

In both cases there is a purely local elastic response to the driving pressure field. Therefore, all our previous results directly apply to these limiting poroelastic cases as well, provided we use the transformations: $\kappa\rightarrow (1-\phi)\kappa$ if $\tau\gg\tau_{\textrm{p}}$, and $\kappa\rightarrow \beta(2\mu+\lambda)\kappa/\phi$ if $\tau\ll\tau_{\textrm{p}}$. 

\section{Conclusions}
Using soft lubrication theory and scaling arguments, we have shown that when a cylinder moves freely close to an elastic or poroelastic wall, the flow-induced pressure field exerts a drag force that resists this motion, but it also deforms the wall, which may in turn increase the gap and reduce this drag, as well as create a supplementary lift. This leads to a complex and rich zoology of inertial motions that link sedimentation, sliding, and spinning, despite the inertialess motion of the fluid. Indeed, it is the wall elasticity combined with the cylinder inertia that are at the origin of all these effects, even at low Reynolds number. The striking observed solutions include non-exhaustively: oscillations, Magnus-like effect, spin-induced reversal, enhanced sedimentation, and long-term steady sliding. While the fully three-dimensional motion of a sphere, or other solid, will have three additional degrees of freedom, we expect many of the qualitative scaling features that we have uncovered to persist.

\appendix
\section{APPENDIX A. Zeroth order: the rigid wall}
In this first Appendix, we detail the derivation of  Eq.~(\ref{tdirect}) at zeroth order in the dimensionless compliance $\kappa$ of the substrate. Equation~(\ref{stokes}) is the Stokes equation for the flow, and the no-slip boundary conditions read: $U(X,Z=0,T)=0$ and $U(X,Z=H,T)=\dot{X}_{\textrm{G}}+\dot{\Theta}$. In addition, the profile of Eq.~(\ref{gap}) becomes: 
\begin{equation}
H(X,T)=\Delta(T)+\left[X-X_{\textrm{G}}(T)\right]^2\ . 
\end{equation}
The corresponding Poiseuille velocity is thus given by:
\begin{equation}
\label{tiltvel}
U=\frac{P_X}2Z\left[Z-\Delta-\left(X-X_{\textrm{G}}\right)^2\right]+\frac{(\dot{X}_{\textrm{G}}+\dot{\Theta})\ Z}{\Delta+\left(X-X_{\textrm{G}}\right)^2}\ . 
\end{equation}

Then, integrating once the volume conservation of Eq.~(\ref{tfe}), with respect to $X$, leads to: 
\begin{equation}
\label{eqspec2}
P_X=\frac{C+12X\dot{\Delta}-12(X-X_{\textrm{G}})^2\dot{X}_{\textrm{G}}+6(\dot{X}_{\textrm{G}}+\dot{\Theta})\left[\Delta+\left(X-X_{\textrm{G}}\right)^2\right]}{\left[\Delta+\left(X-X_{\textrm{G}}\right)^2\right]^3}\ ,
\end{equation}
where $C(T)=-\left(8 \Delta \dot{\Theta} + 4 \Delta \dot{X}_{\textrm{G}} + 12 \dot{\Delta} X_{\textrm{G}}\right)$ is an integration constant, that was identified thanks to the assumed vanishing lubrication pressure $P$ at $X=\pm\infty$. In this case, a second spatial integration leads to:
\begin{equation}
\label{tiltpres}
P=-\frac{3\dot{\Delta}+2(\dot{\Theta}-\dot{X}_{\textrm{G}})(X-X_{\textrm{G}})}{\left[\Delta+\left(X-X_{\textrm{G}}\right)^2\right]^2}\ .
\end{equation}

The pressure is not an even function in $X$ due to the transverse motion, and therefore there is a tangential drag associated to it, in addition to the normal one. We use Eq.~(\ref{dragp}) to evaluate both projections. By parity, the total dimensionless pressure-induced drag force along $Z$ is thus:
\begin{equation}
\label{zeroperp}
D_{p,\perp}=\int_{-\infty}^{\infty}dX\ P=-\frac{3\pi}{2}\frac{\dot{\Delta}}{\Delta^{3/2}}\ .
\end{equation}
Similarly, the total dimensionless pressure-induced drag force along $X$ reads:
\begin{equation}
\label{zeropara}
D_{p,\parallel}=-\sqrt{2\epsilon}\int_{-\infty}^{\infty}dX\ (X-X_{\textrm{G}})\ P=\pi\sqrt{2\epsilon}\ \frac{\dot{\Theta}-\dot{X}_{\textrm{G}}}{\sqrt{\Delta}}\ ,
\end{equation}
which is smaller in magnitude -- by a factor $\sim\sqrt{\epsilon}\ll1$ -- than the orthogonal one along $Z$. 

It is important to highlight that we had to go to the next order in $\sqrt{\epsilon}$ to obtain the pressure-induced drag force $D_{p,\parallel}$ in the tangential direction, which is now of comparable magnitude as the tangential drag $D_{\sigma,\parallel}$ obtained from the dominant viscous stress component: $\sigma_{zx}\approx\eta \partial_zu$. Therefore, one has to calculate the latter through Eq.~(\ref{drags}) with $\kappa=0$:
\begin{equation}
D_{\sigma,\parallel}=-\sqrt{\frac{\epsilon}{2}}\int_{-\infty}^{\infty}dX\ U_Z|_{Z=H}\ .
\end{equation}
Using Eqs.~(\ref{tiltvel}) and~(\ref{tiltpres}), it becomes:
\begin{equation}
\label{zerosig}
D_{\sigma,\parallel}=-\pi\sqrt{2\epsilon}\frac{\dot{\Theta}}{\sqrt{\Delta}}\ ,
\end{equation}
which precisely compensates the part of $D_{p,\parallel}$ that depends on $\dot{\Theta}$. 

Knowing the dominant drag in each direction, one can now study the motion of the cylinder in the presence of gravity and buoyancy. The $Z$-projection of the balance of linear momentum reads:
\begin{equation}
\ddot{\Delta}+\xi\frac{\dot{\Delta}}{\Delta^{3/2}}+\cos\alpha=0\ .
\end{equation}
Thus, the sedimentation motion is decoupled from the others.  In contrast, the sliding motion is coupled to the sedimentation motion through the $X$-projection of the balance of linear momentum, as given by:
\begin{equation}
\ddot{X}_{\textrm{G}}+\frac{2\epsilon\xi}{3}\ \frac{\dot{X}_{\textrm{G}}}{\sqrt{\Delta}}-\sqrt{\frac{\epsilon}{2}}\sin\alpha=0\ .
\end{equation}
Finally, the spinning motion can be obtained by the balance of angular momentum that reads:
\begin{equation}
\frac{mr^2}{2}\ddot{\theta}=r\ d_{\sigma,\parallel}\ ,
\end{equation}
where the pressure-induced torque is zero since the pressure-induced force acts along a radius of the cylinder. This can be non-dimensionalized as:
\begin{equation}
\ddot{\Theta}+\frac{4\epsilon\xi}{3}\frac{\dot{\Theta}}{\sqrt{\Delta}}=0\ ,
\end{equation} 
which results in the trivial non-spinning solution, if $\dot{\Theta}(0)=0$, due to the absence of driving force. This statement is modified for a soft wall, as studied below.
 
\section{APPENDIX B. First-order correction: the soft compressible wall}
Here, we detail the derivation of the central Eq.~(\ref{tdirect}) at first order in the dimensionless compliance $\kappa$ of the substrate. Solving Eq.~(\ref{stokes}) with the new boundary conditions: $U(X,Z=-\kappa P,T)=0$, and $U(X,Z=H-\kappa P,T)=\dot{X}_{\textrm{G}}+\dot{\Theta}$, and the gap profile of Eq.~(\ref{gap}), and conserving the volume of the fluid through Eq.~(\ref{tfe}), leads to Eq.~(\ref{eqgen3}). Since $P(X,T)$ depends on $X$, a direct spatial integration of this equation would lead to an integro-differential equation. We restrict ourselves to perturbation theory in $\kappa\ll1$, consistent with the assumption of linear elasticity:
\begin{eqnarray}
P&=&P^{(0)}+\kappa P^{(1)}\ ,\\
D_{p,\perp}&=&D_{p,\perp}^{(0)}+\kappa D_{p,\perp}^{(1)}\ ,\\
D_{p,\parallel}&=&D_{p,\parallel}^{(0)}+\kappa D_{p,\parallel}^{(1)}\ ,\\
D_{\sigma,\parallel}&=&D_{\sigma,\parallel}^{(0)}+\kappa D_{\sigma,\parallel}^{(1)}\ ,
\end{eqnarray}
where both $P^{(0)}$ and $P^{(1)}$ are assumed to vanish at infinity. 

Equation~(\ref{eqgen3}) at zeroth order in $\kappa$ is equivalent to Eq.~(\ref{eqspec2}), so that the zeroth order pressure follows from Eq.~(\ref{tiltpres}): 
\begin{equation}
P^{(0)}=-\frac{3\dot{\Delta}+2(\dot{\Theta}-\dot{X}_{\textrm{G}})(X-X_{\textrm{G}})}{\left[\Delta+\left(X-X_{\textrm{G}}\right)^2\right]^2}\ ,
\end{equation}
while the corresponding zeroth order drag forces from Eqs.~(\ref{zeroperp}), (\ref{zeropara}), and~(\ref{zerosig}) are:
\begin{eqnarray}
D_{p,\perp}^{(0)}&=&-\frac{3\pi}{2}\frac{\dot{\Delta}}{\Delta^{3/2}}\ ,\\
D_{p,\parallel}^{(0)}&=&\pi\sqrt{2\epsilon}\ \frac{\dot{\Theta}-\dot{X}_{\textrm{G}}}{\sqrt{\Delta}}\ ,\\
D_{\sigma,\parallel}^{(0)}&=&-\pi\sqrt{2\epsilon}\frac{\dot{\Theta}}{\sqrt{\Delta}}\ .
\end{eqnarray}

Expressing Eq.~(\ref{eqgen3}) at first order in $\kappa$ then yields:
\begin{equation}
\left[\left(\Delta+(X-X_{\textrm{G}})^2\right)^3P^{(1)}_X+3\left(\Delta+(X-X_{\textrm{G}})^2\right)^2P^{(0)}P^{(0)}_X-6(\dot{\Theta}+\dot{X}_{\textrm{G}})P^{(0)}\right]_X=12P^{(0)}_T\ .
\end{equation}
Proceeding as in Appendix A, using three spatial integrations and the above-mentioned boundary conditions, one gets the normal and tangential pressure-induced drag forces as:
\begin{eqnarray}
D_{p,\perp}^{(1)}&=&\frac{45\pi\ddot{\Delta}}{16\Delta^{7/2}}-\frac{63\pi\dot{\Delta}^2}{8\Delta^{9/2}}+\frac{3\pi(\dot{\Theta}-\dot{X}_{\textrm{G}})^2}{8\Delta^{7/2}}\ ,\\
D_{p,\parallel}^{(1)}&=&\sqrt{\frac{\epsilon}{2}}\left[ \frac{23\pi\dot{\Delta}(\dot{\Theta}-\dot{X}_{\textrm{G}})}{8\Delta^{7/2}}+\frac{\pi(\ddot{X}_{\textrm{G}}-\ddot{\Theta})}{2\Delta^{5/2}}\right]\ ,
\end{eqnarray}
that are consistent with the steady-state results ~\cite[]{Skotheim2004,Skotheim2005} when $\Delta=\dot{X}_{\textrm{G}}\equiv1$ and $\Theta\equiv0$. 

In order to calculate the remaining first order viscous stress, one expresses the velocity gradient at the surface of the cylinder:
\begin{equation}
U_Z|_{Z=H-\kappa P}=U_Z|_{Z=H-\kappa P}^{(0)}+\kappa U_Z|_{Z=H-\kappa P}^{(1)}\ ,
\end{equation}
where:
\begin{equation}
U_Z|_{Z=H-\kappa P}^{(1)}=\frac{P_X^{(1)}}{2}\left[\Delta+(X-X_{\textrm{G}})^2\right]+\frac{P^{(0)}P^{(0)}_X}{2}-\frac{(\dot{\Theta}+\dot{X}_{\textrm{G}})P^{(0)}}{\left[\Delta+(X-X_{\textrm{G}})^2\right]^2}\ .
\end{equation}
Therefore, using Eq.~(\ref{drags}), one gets:
\begin{equation}
D_{\sigma,\parallel}^{(1)}=\sqrt{\frac{\epsilon}{2}}\left[\frac{\pi(\ddot{\Theta}-\ddot{X}_{\textrm{G}})}{4\Delta^{5/2}}+\frac{\pi\dot{\Delta}\dot{X}_{\textrm{G}}}{2\Delta^{7/2}}-\frac{19\pi\dot{\Delta}\dot{\Theta}}{8\Delta^{7/2}}\right]\ .
\end{equation}

Finally, the balance of linear and angular momentum leads to the general coupled system of three equations:
\begin{eqnarray}
\label{zdirect}
\ddot{\Delta}+\xi\frac{\dot{\Delta}}{\Delta^{3/2}}+\frac{\kappa\xi}{4}\left[21\frac{\dot{\Delta}^2}{\Delta^{9/2}}-\frac{(\dot{\Theta}-\dot{X}_{\textrm{G}})^2}{\Delta^{7/2}}-\frac{15}{2}\frac{\ddot{\Delta}}{\Delta^{7/2}}\right]+\cos\alpha&=&0\ ,\\ 
\label{xdirect}
\ddot{X}_{\textrm{G}}+\frac{2\epsilon\xi}{3}\ \frac{\dot{X}_{\textrm{G}}}{\sqrt{\Delta}}+\frac{\kappa\epsilon\xi}{6}\left[\frac{19}{4}\frac{\dot{\Delta}\dot{X}_{\textrm{G}}}{\Delta^{7/2}}-\frac{\dot{\Delta}\dot{\Theta}}{\Delta^{7/2}}+\frac12\frac{\ddot{\Theta}-\ddot{X}_{\textrm{G}}}{\Delta^{5/2}}\right]-\sqrt{\frac{\epsilon}{2}}\sin\alpha&=&0\ ,\\
\ddot{\Theta}+\frac{4\epsilon\xi}{3}\frac{\dot{\Theta}}{\sqrt{\Delta}}+\frac{\kappa\epsilon\xi}{3}\left[\frac{19}{4}\frac{\dot{\Delta}\dot{\Theta}}{\Delta^{7/2}}-\frac{\dot{\Delta}\dot{X}_{\textrm{G}}}{\Delta^{7/2}}+\frac12\frac{\ddot{X}_{\textrm{G}}-\ddot{\Theta}}{\Delta^{5/2}}\right]&=&0\ ,
\end{eqnarray}
which corresponds to Eq.~(\ref{tdirect}).

\bibliographystyle{jfm}

\begin{thebibliography}{38}
\expandafter\ifx\csname natexlab\endcsname\relax\def\natexlab#1{#1}\fi
\expandafter\ifx\csname bibnamefont\endcsname\relax
\def\bibnamefont#1{#1}\fi
\expandafter\ifx\csname bibfnamefont\endcsname\relax
\def\bibfnamefont#1{#1}\fi
\expandafter\ifx\csname citenamefont\endcsname\relax
\def\cite[]namefont#1{#1}\fi
\expandafter\ifx\csname url\endcsname\relax
\def\url#1{\texttt{#1}}\fi
\expandafter\ifx\csname urlprefix\endcsname\relax\def\urlprefix{URL }\fi
\providecommand{\bibinfo}[2]{#2}
\providecommand{\eprint}[2][]{\url{#2}}

\bibitem[Balmforth et al.(2010)]{Balmforth2010}
\bibinfo{author}{\bibfnamefont{N. J.}~\bibnamefont{Balmforth}},
\bibinfo{author}{\bibfnamefont{C. J.}~\bibnamefont{Cawthorn}}, \bibnamefont{and}
\bibinfo{author}{\bibfnamefont{R. V.}~\bibnamefont{Craster}},
\bibinfo{journal}{J. Fluid Mech.} \textbf{\bibinfo{volume}{646}},
\bibinfo{pages}{339} (\bibinfo{year}{2010}).

\bibitem[Batchelor(1967)]{Batchelor1967}
\bibinfo{author}{\bibfnamefont{G. K.}~\bibnamefont{Batchelor}},
\bibinfo{title}{An Introduction to Fluid Dynamics},
\bibinfo{publisher}{Cambridge University Press, Cambridge, England} (\bibinfo{year}{1967}).

\bibitem[Biot(1941)]{Biot1941}
\bibinfo{author}{\bibfnamefont{M. A.}~\bibnamefont{Biot}},
\bibinfo{journal}{J. Appl. Phys.} \textbf{\bibinfo{volume}{12}},
\bibinfo{pages}{155} (\bibinfo{year}{1941}).

\bibitem[Brenner(1962)]{Brenner1962}
\bibinfo{author}{\bibfnamefont{H.}~\bibnamefont{Brenner}},
\bibinfo{journal}{J. Fluid Mech.} \textbf{\bibinfo{volume}{12}},
\bibinfo{pages}{35} (\bibinfo{year}{1962}).

\bibitem[Burridge and Keller(1981)]{Burridge1981}
\bibinfo{author}{\bibfnamefont{R.}~\bibnamefont{Burridge}}, \bibnamefont{and}
\bibinfo{author}{\bibfnamefont{J. B.}~\bibnamefont{Keller}},
\bibinfo{journal}{J. Acoust. Soc. Am.} \textbf{\bibinfo{volume}{70}},
\bibinfo{pages}{1140} (\bibinfo{year}{1981}).

\bibitem[Byun et al.(2013)]{Byun2013}
\bibinfo{author}{\bibfnamefont{S.}~\bibnamefont{Byun}},
\bibinfo{author}{\bibfnamefont{S.}~\bibnamefont{Son}},
\bibinfo{author}{\bibfnamefont{D.}~\bibnamefont{Amodei}},
\bibinfo{author}{\bibfnamefont{N.}~\bibnamefont{Cermak}},
\bibinfo{author}{\bibfnamefont{J.}~\bibnamefont{Shaw}},
\bibinfo{author}{\bibfnamefont{J. H.}~\bibnamefont{Kang}},
\bibinfo{author}{\bibfnamefont{V. C.}~\bibnamefont{Hecht}},
\bibinfo{author}{\bibfnamefont{M.}~\bibnamefont{Winslow}},
\bibinfo{author}{\bibfnamefont{T.}~\bibnamefont{Jacks}},
\bibinfo{author}{\bibfnamefont{P.}~\bibnamefont{Mallick}}, \bibnamefont{and}
\bibinfo{author}{\bibfnamefont{S. R.}~\bibnamefont{Manalis}},
\bibinfo{journal}{PNAS (USA)} \textbf{\bibinfo{volume}{110}},
\bibinfo{pages}{7580} (\bibinfo{year}{2013}).

\bibitem[Campbell(1989)]{Campbell1989}
\bibinfo{author}{\bibfnamefont{C. S.}~\bibnamefont{Campbell}},
\bibinfo{journal}{J. Geol.} \textbf{\bibinfo{volume}{97}},
\bibinfo{pages}{653} (\bibinfo{year}{1989}).

\bibitem[Cawthorn and Balmforth(2010)]{Cawthorn2010}
\bibinfo{author}{\bibfnamefont{C. J.}~\bibnamefont{Cawthorn}}, \bibnamefont{and}
\bibinfo{author}{\bibfnamefont{N. J.}~\bibnamefont{Balmforth}},
\bibinfo{journal}{J. Fluid Mech.} \textbf{\bibinfo{volume}{646}},
\bibinfo{pages}{327} (\bibinfo{year}{2010}).

\bibitem[Davis et al.(1986)]{Davis1986}
\bibinfo{author}{\bibfnamefont{R. H.}~\bibnamefont{Davis}},
\bibinfo{author}{\bibfnamefont{J.-M.}~\bibnamefont{Serayssol}}, \bibnamefont{and}
\bibinfo{author}{\bibfnamefont{E. J.}~\bibnamefont{Hinch}},
\bibinfo{journal}{Phys. Fluids} \textbf{\bibinfo{volume}{163}},
\bibinfo{pages}{479} (\bibinfo{year}{1986}).

\bibitem[Dupeux et al.(2011)]{Dupeux2011}
\bibinfo{author}{\bibfnamefont{G.}~\bibnamefont{Dupeux}},
\bibinfo{author}{\bibfnamefont{C.}~\bibnamefont{Cohen}},
\bibinfo{author}{\bibfnamefont{A.}~\bibnamefont{Le Goff}},
\bibinfo{author}{\bibfnamefont{D.}~\bibnamefont{Qu\'er\'e}}, \bibnamefont{and}
\bibinfo{author}{\bibfnamefont{C.}~\bibnamefont{Clanet}},
\bibinfo{journal}{J. Fluids and Structures} \textbf{\bibinfo{volume}{27}},
\bibinfo{pages}{659} (\bibinfo{year}{2011}).

\bibitem[Glenne et al.(1987)]{Glenne1987}
\bibinfo{author}{\bibfnamefont{B.}~\bibnamefont{Glenne}},
\bibinfo{journal}{J. Tribology} \textbf{\bibinfo{volume}{109}},
\bibinfo{pages}{614} (\bibinfo{year}{1987}).

\bibitem[Goldman et al.(1967)]{Goldman1967}
\bibinfo{author}{\bibfnamefont{A. J.}~\bibnamefont{Goldman}},
\bibinfo{author}{\bibfnamefont{R. G.}~\bibnamefont{Cox}}, \bibnamefont{and}
\bibinfo{author}{\bibfnamefont{H.}~\bibnamefont{Brenner}},
\bibinfo{journal}{Chem. Eng. Sci.} \textbf{\bibinfo{volume}{22}},
\bibinfo{pages}{637} (\bibinfo{year}{1967}).

\bibitem[Goldman et al.(1967)]{Goldman1967b}
\bibinfo{author}{\bibfnamefont{A. J.}~\bibnamefont{Goldman}},
\bibinfo{author}{\bibfnamefont{R. G.}~\bibnamefont{Cox}}, \bibnamefont{and}
\bibinfo{author}{\bibfnamefont{H.}~\bibnamefont{Brenner}},
\bibinfo{journal}{Chem. Eng. Sci.} \textbf{\bibinfo{volume}{22}},
\bibinfo{pages}{653} (\bibinfo{year}{1967}).

\bibitem[Goldsmith(1971)]{Goldsmith1971}
\bibinfo{author}{\bibfnamefont{H. L.}~\bibnamefont{Goldsmith}},
\bibinfo{journal}{Fed Proc.} \textbf{\bibinfo{volume}{30}},
\bibinfo{pages}{1578} (\bibinfo{year}{1971}).

\bibitem[Gondret et al.(1999)]{Gondret1999}
\bibinfo{author}{\bibfnamefont{P.}~\bibnamefont{Gondret}},
\bibinfo{author}{\bibfnamefont{E.}~\bibnamefont{Hallouin}},
\bibinfo{author}{\bibfnamefont{M.}~\bibnamefont{Lance}}, \bibnamefont{and}
\bibinfo{author}{\bibfnamefont{L.}~\bibnamefont{Petit}},
\bibinfo{journal}{Phys. Fluids} \textbf{\bibinfo{volume}{11}},
\bibinfo{pages}{2803} (\bibinfo{year}{1999}).

\bibitem[Gopinath and Mahadevan(2011)]{Gopinath2011}
\bibinfo{author}{\bibfnamefont{A.}~\bibnamefont{Gopinath}}, \bibnamefont{and}
\bibinfo{author}{\bibfnamefont{L.}~\bibnamefont{Mahadevan}},
\bibinfo{journal}{Proc. R. Soc. Lond. Ser. A} \textbf{\bibinfo{volume}{467}},
\bibinfo{pages}{1665} (\bibinfo{year}{2011}).

\bibitem[Grodzinsky et al.(1978)]{Grodzinsky1978}
\bibinfo{author}{\bibfnamefont{A. J.}~\bibnamefont{Grodzinsky}},
\bibinfo{author}{\bibfnamefont{H.}~\bibnamefont{Lipshitz}}, \bibnamefont{and}
\bibinfo{author}{\bibfnamefont{M. J.}~\bibnamefont{Glimcher}},
\bibinfo{journal}{Nature} \textbf{\bibinfo{volume}{275}},
\bibinfo{pages}{448} (\bibinfo{year}{1978}).

\bibitem[Hocking(1973)]{Hocking1973}
\bibinfo{author}{\bibfnamefont{L. M.}~\bibnamefont{Hocking}},
\bibinfo{journal}{J. Engg. Math.} \textbf{\bibinfo{volume}{7}},
\bibinfo{pages}{207} (\bibinfo{year}{1973}).

\bibitem[Jeffrey and Onishi(1981)]{Jeffrey1981}
\bibinfo{author}{\bibfnamefont{D. J.}~\bibnamefont{Jeffrey}}, \bibnamefont{and}
\bibinfo{author}{\bibfnamefont{Y.}~\bibnamefont{Onishi}},
\bibinfo{journal}{Quart. J. Mech. App. Math.} \textbf{\bibinfo{volume}{34}},
\bibinfo{pages}{129} (\bibinfo{year}{1981}).

\bibitem[Johnson(1985)]{Johnson1985}
\bibinfo{author}{\bibfnamefont{K. L.}~\bibnamefont{Johnson}},
\bibinfo{title}{Contact Mechanics},
\bibinfo{publisher}{Cambridge University Press, Cambridge, England} (\bibinfo{year}{1985}).

\bibitem[Ma et al.(2003)]{Ma2003}
\bibinfo{author}{\bibfnamefont{K.-F.}~\bibnamefont{Ma}},
\bibinfo{author}{\bibfnamefont{E. E.}~\bibnamefont{Brodsky}},
\bibinfo{author}{\bibfnamefont{J.}~\bibnamefont{Mori}},
\bibinfo{author}{\bibfnamefont{C.}~\bibnamefont{Ji}},
\bibinfo{author}{\bibfnamefont{T.-R. A.}~\bibnamefont{Song}}, \bibnamefont{and}
\bibinfo{author}{\bibfnamefont{H.}~\bibnamefont{Kanamori}},
\bibinfo{journal}{Geophys. Res. Lett.} \textbf{\bibinfo{volume}{30}},
\bibinfo{pages}{1244} (\bibinfo{year}{2003}).

\bibitem[Mani et al.(2012)]{Mani2012}
\bibinfo{author}{\bibfnamefont{M.}~\bibnamefont{Mani}},
\bibinfo{author}{\bibfnamefont{A.}~\bibnamefont{Gopinath}}, \bibnamefont{and}
\bibinfo{author}{\bibfnamefont{L.}~\bibnamefont{Mahadevan}},
\bibinfo{journal}{Phys. Rev. Lett.} \textbf{\bibinfo{volume}{226104}},
\bibinfo{pages}{108} (\bibinfo{year}{2012}).

\bibitem[Mow et al.(1984)]{Mow1984}
\bibinfo{author}{\bibfnamefont{V. C.}~\bibnamefont{Mow}},
\bibinfo{author}{\bibfnamefont{M. H.}~\bibnamefont{Holmes}}, \bibnamefont{and}
\bibinfo{author}{\bibfnamefont{W. M.}~\bibnamefont{Lai}},
\bibinfo{journal}{J. Biomech.} \textbf{\bibinfo{volume}{17}},
\bibinfo{pages}{377} (\bibinfo{year}{1984}).

\bibitem[Mow et al.(2002)]{Mow2002}
\bibinfo{author}{\bibfnamefont{V. C.}~\bibnamefont{Mow}}, \bibnamefont{and}
\bibinfo{author}{\bibfnamefont{X. E.}~\bibnamefont{Guo}},
\bibinfo{journal}{Annu. Rev. Biomed. Eng.} \textbf{\bibinfo{volume}{4}},
\bibinfo{pages}{175} (\bibinfo{year}{2002}).

\bibitem[Oron et al.(1997)]{Oron1997}
\bibinfo{author}{\bibfnamefont{A.}~\bibnamefont{Oron}},
\bibinfo{author}{\bibfnamefont{S.}~\bibnamefont{Davis}}, \bibnamefont{and}
\bibinfo{author}{\bibfnamefont{S.}~\bibnamefont{Bankoff}},
\bibinfo{journal}{Rev. Mod. Phys.} \textbf{\bibinfo{volume}{69}},
\bibinfo{pages}{931} (\bibinfo{year}{1997}).

\bibitem[Reynolds(1886)]{Reynolds1886}
\bibinfo{author}{\bibfnamefont{O.}~\bibnamefont{Reynolds}},
\bibinfo{journal}{Philos. Trans. R. Soc. Lond.} \textbf{\bibinfo{volume}{177}},
\bibinfo{pages}{157} (\bibinfo{year}{1886}).

\bibitem[Sekimoto and Leibler(1993)]{Sekimoto1993}
\bibinfo{author}{\bibfnamefont{K.}~\bibnamefont{Sekimoto}}, \bibnamefont{and}
\bibinfo{author}{\bibfnamefont{L.}~\bibnamefont{Leibler}},
\bibinfo{journal}{Europhys. Lett.} \textbf{\bibinfo{volume}{23}},
\bibinfo{pages}{113} (\bibinfo{year}{1993}).

\bibitem[Sekimoto and Rabin(1994)]{Sekimoto1994}
\bibinfo{author}{\bibfnamefont{K.}~\bibnamefont{Sekimoto}}, \bibnamefont{and}
\bibinfo{author}{\bibfnamefont{Y.}~\bibnamefont{Rabin}},
\bibinfo{journal}{Europhys. Lett.} \textbf{\bibinfo{volume}{27}},
\bibinfo{pages}{445} (\bibinfo{year}{1994}).

\bibitem[Skotheim and Mahadevan(2004)]{Skotheim2004}
\bibinfo{author}{\bibfnamefont{J. M.}~\bibnamefont{Skotheim}}, \bibnamefont{and}
\bibinfo{author}{\bibfnamefont{L.}~\bibnamefont{Mahadevan}},
\bibinfo{journal}{Phys. Rev. Lett.} \textbf{\bibinfo{volume}{92}},
\bibinfo{pages}{245509} (\bibinfo{year}{2004}).

\bibitem[Skotheim and Mahadevan(2004)]{Skotheim2004b}
\bibinfo{author}{\bibfnamefont{J. M.}~\bibnamefont{Skotheim}}, \bibnamefont{and}
\bibinfo{author}{\bibfnamefont{L.}~\bibnamefont{Mahadevan}},
\bibinfo{journal}{Proc. R. Soc. Lond. Ser. A} \textbf{\bibinfo{volume}{460}},
\bibinfo{pages}{1995} (\bibinfo{year}{2004}).

\bibitem[Skotheim and Mahadevan(2005)]{Skotheim2005}
\bibinfo{author}{\bibfnamefont{J. M.}~\bibnamefont{Skotheim}}, \bibnamefont{and}
\bibinfo{author}{\bibfnamefont{L.}~\bibnamefont{Mahadevan}},
\bibinfo{journal}{Phys. Fluids} \textbf{\bibinfo{volume}{17}},
\bibinfo{pages}{092101} (\bibinfo{year}{2005}).

\bibitem[Snoeijer et al.(2013)]{Snoeijer2013}
\bibinfo{author}{\bibfnamefont{J.}~\bibnamefont{Snoeijer}},
\bibinfo{author}{\bibfnamefont{J.}~\bibnamefont{Eggers}}, \bibnamefont{and}
\bibinfo{author}{\bibfnamefont{C. H.}~\bibnamefont{Venner}},
\bibinfo{journal}{Phys. Fluids} \textbf{\bibinfo{volume}{25}},
\bibinfo{pages}{101705} (\bibinfo{year}{2013}).

\bibitem[Trahan and Hussey(1985)]{Trahan1985}
\bibinfo{author}{\bibfnamefont{J. F.}~\bibnamefont{Trahan}}, \bibnamefont{and}
\bibinfo{author}{\bibfnamefont{R. G.}~\bibnamefont{Hussey}},
\bibinfo{journal}{Phys. Fluids} \textbf{\bibinfo{volume}{28}},
\bibinfo{pages}{2961} (\bibinfo{year}{1985}).

\bibitem[Weekley et al.(2006)]{Weekley2006}
\bibinfo{author}{\bibfnamefont{S. J.}~\bibnamefont{Weekley}},
\bibinfo{author}{\bibfnamefont{S. L.}~\bibnamefont{Waters}}, \bibnamefont{and}
\bibinfo{author}{\bibfnamefont{O. E.}~\bibnamefont{Jensen}},
\bibinfo{journal}{Q. J. Mech. Appl. Math.} \textbf{\bibinfo{volume}{59}},
\bibinfo{pages}{277} (\bibinfo{year}{2006}).

\bibitem[Wehbeh et al.(1993)]{Wehbeh1993}
\bibinfo{author}{\bibfnamefont{E. G.}~\bibnamefont{Wehbeh}},
\bibinfo{author}{\bibfnamefont{T. J.}~\bibnamefont{Ui}}, \bibnamefont{and}
\bibinfo{author}{\bibfnamefont{R. G.}~\bibnamefont{Hussey}},
\bibinfo{journal}{Phys. Fluids} \textbf{\bibinfo{volume}{5}},
\bibinfo{pages}{25} (\bibinfo{year}{1993}).

\end{thebibliography}

\end{document}